\documentclass{article}
\pdfoutput=1

\usepackage{PRIMEarxiv}

\usepackage[utf8]{inputenc} 
\usepackage[T1]{fontenc}    
\usepackage{hyperref}       
\usepackage{comment}
\usepackage{url}    
\usepackage{longtable}
\usepackage{booktabs}       
\usepackage{amsfonts}       
\usepackage{nicefrac}       
\usepackage{microtype}      
\usepackage{lipsum}
\usepackage{fancyhdr}       
\usepackage{graphicx}       
\graphicspath{{media/}}     
\usepackage{amssymb,amsmath}
\usepackage{amsthm}
\newtheorem{lemma}{Lemma}
\numberwithin{equation}{section}
\theoremstyle{definition}
\newtheorem{theorem}{Theorem}[section]

\newtheorem{example}{Example}[section]

\title{A Tutorial on Asymptotic Properties of Statistical Estimators with Applications to COVID-19 Data
}

\author{
  Elvis Han Cui \\
  PhD Candidate in Biostatistics,\\
  University of California, Los Angeles\\
  elviscuihan@g.ucla.edu} 

\begin{document}
\maketitle

\begin{abstract}
Asymptotic properties of statistical estimators play a significant role  both 
in practice and in theory.  However, many asymptotic results in statistics rely heavily on the independent and identically distributed (iid) assumption, which is not realistic when we have fixed designs. In this article, we build a roadmap of general procedures for deriving asymptotic properties under fixed designs and the observations need not to be iid. We further provide their applications in many statistical applications. Finally, we apply our results to Poisson regression using a COVID-19 dataset as an illustration to demonstrate the power of these results in practice.
\end{abstract}



\section{Introduction}

A good estimator should, at least in the asymptotic sense, be close to the true quantity that it wishes to estimate and we should be able to give uncertainty measure based on a finite sample size. An estimator with well-behaved asymptotic properties can help clinicians in many ways such as reducing the number of patients needed in a trial, cutting down the budget for toxicology studies and providing insightful findings for late phase trials. Suggested by Sr. Fisher  \cite{fisher1925theory}, generations of statisticians have worked on the so-called "consistency" and "asymptotic normality" of estimators. The former is based on different versions of law of large numbers (LLN) and the later is based on various types of central limit theorems (CLT) \cite{chung2001course}. In addition to these two main tools, statisticians also apply other important but less well-known results in probability theory and other mathematical fields. To name a few, extreme value theory for distributions of maxima and minima \cite{gnedenko1954limit}, convex analysis for checking the optimality of a statistical design \cite{kiefer1974general}, asymptotic relative efficiency (ARE) of an estimator \cite{bickel1993efficient}, concentration inequalities for finite sample properties and selection consistency \cite{vapnik2015uniform} and other non-normal limits , robustness and simultaneous confidence bands of common statistical estimators \cite{huber2011robust, andersen2012statistical}.

Despite of different properties, consistency and asymptotic normality are still the most celebrated and important properties of statistical estimators in either academia or industry. Hence, in the following, we present a roadmap to consistency and asymptotic normality. Then we provide their applications in toxicology studies and clinical trials using a COVID-19 dataset.

\section{Consistency}

The weak and strong consistency are based on weak law of large numbers (WLLN) and strong law of large numbers (SLLN), respectively. The naive example is the consistency of sample mean: we measure the height of a subject multiple times and then take the average of all measurements.  The theoretical foundation is based on \emph{Chebyshev's inequality} and is known as the \emph{Khinchin's WLLN}.

\subsection{Weak Consistency}

\begin{theorem}[Khinchin's WLLN]
	Suppose that $X_1,\cdots,X_n$ are independent and identically distributed (iid) with finite means and variances, $var(X_i)=\sigma^2$. Then
	\begin{align}
		\mathbb{P}(\frac{1}{n}|\sum_{i=1}^nX_i|>\epsilon)\le \frac{\sigma^2}{n\epsilon^2}\rightarrow 0
	\end{align}
	as $n$ goes to  infinity where $\epsilon$ is a small positive number.
\end{theorem}

A refinement of Khinchin's LLN is based on the Hoeffding's inequality \cite{hoeffding1994probability}, which provides the finite sample bounds of the estimator $\bar{X}_n=\sum_{i=1}^nX_i$.

\begin{example}[Hoeffding's inequality for boosting algorithm]
	Consider a binary classification problem and an algorithm for solving it. Suppose the algorithm has probability $\frac{1}{2}+\delta$ of correctly classifying a sample where $\delta > 0$ is small. For any $\epsilon\in(0,1)$, if we run the algorithm $n$ times where 
	$$n\ge \frac{1}{2\delta^2}\ln(\frac{1}{\epsilon}),$$
	 and take the majority as the final decision, then with probability greater than $1-\epsilon$, we can correctly classify the sample. Indeed, let $X_i,i=1,2,\cdots,n$ be a sequence of iid Bernoulli variables with success probability $\frac{1}{2}+\epsilon$, then
	$$\mathbb{P}(\bar{X}-\frac{1}{2}<0)\le \exp(-2n\delta^2)\le \epsilon.$$
	Hence, $\mathbb{P}(\bar{X}-\frac{1}{2}\ge0)\ge 1-\epsilon$ but the left side is the probability of making a correct classification.
\end{example}

However, the Khincin's LLN and Hoeffding's inequality are rarely used in clinical trials for the following reason. Different patients have different demographic characteristics such as age, gender, weight and height, hence, it is almost unrealistic to assume the measurements are iid among patients.  Further, in a longitudinal study, the repeated measurements are not even independent within a subject. Thus, extensions of LLN are considered and mathematicians have developed a large group of available tools. Before we present the general results, two natural extensions are immediate.

\begin{example}[Independent but not identically distributed sequence] Let $X_i,i=1,2,\cdots$ be a sequence of independent variables with mean $\mu$ and variances $\sigma_i^2,i=1,2,\cdots$. Let $w_i=\frac{1}{\sigma_i^2}$ and suppose that  $\sum_i^\infty w_i=+\infty$, then the weighted sample average is weakly consistent, i.e., 
	$$\sum_{i=1}^n\frac{w_i}{\sum_{i=1}^nw_i}X_i\rightarrow_p\mu.$$
	
	This is a simple application of Chebyshev's inequality:
	\begin{align}
		\mathbb{P}(|\sum_{i=1}^n\frac{w_i}{\sum_{i=1}^nw_i}X_i-\mu|>\epsilon)&\le\frac{1}{\epsilon^2}\sum_{i=1}^n\left(\frac{w_i^2}{(\sum_{i=1}^nw_i)^2}var(X_i)\right)\nonumber\\
		&=\frac{1}{\epsilon^2(\sum_{i=1}^nw_i)},
	\end{align}
which goes to 0 as $n$ goes to infinity.

\end{example}

\begin{example}[Weak consistency of dependent variables]
	Let $X_1,X_2,\cdots$ be a sequence of mean zero variables with variances $\sigma_i^2$ bounded by $C$ (i.e., $\sup_i\sigma_i^2\le C$). Suppose $Cov(X_n,X_m)=\rho(n-m)$ for $n>m$ and $\rho(n)\rightarrow 0$. Then we have
	$$\frac{1}{n}\sum_{i=1}^nX_i\rightarrow_p0.$$
	Indeed, by Chebyshev's inequality and assume $\rho(0)=C$,
	\begin{align*}
		\mathbb{P}(|\frac{1}{n}\sum_{i=1}^nX_i|>\epsilon)&\le\frac{1}{\epsilon^2}\frac{1}{n^2}\sum_{i,j}^n|\rho(|i-j|)|\\
		&\le\frac{2}{n\epsilon^2}\sum_{k=0}^{n-1}\frac{n-k}{n}|\rho(k)|\\
		&\le\frac{2}{n\epsilon^2}\sum_{k=0}^{n-1}|\rho(k)|,
	\end{align*}
then by Cesaro's mean lemma, we have $\lim_n\frac{1}{n}\sum_{k=0}^{n-1}|\rho(k)|=\lim_k|\rho(k)|=0$.
	
\end{example}
The definitive answer of WLLN is given by William Feller and Andrey Kolmogorov.

\begin{theorem}[Feller-Kolmogorov WLLN \cite{feller1967introduction, dabrowska2019}]
	Let $\mathbf{X}^{(n)}=\{X_{n,j}:j=1,\cdots,r_n\}\},n\ge 1$ be a triangular array with independent components and adapted to the filtration $\mathbf{F}^{(n)}$. Set $S_n=\sum_{j=1}^{r_n}X_{n,j}$ and let $v_n$ be a positive deterministic sequence such that $v_n\rightarrow\infty$. Put
	$$Y_{n,j}=X_{n,j}\mathbb{I}(|X_{n,j}|\le v_n)\text{ and }B_n=\sum_{j=1}^{r_n}\mathbb{E}(Y_{n,j}|\mathcal{F}_{n,j-1})$$
	and if 
	$$\sum_{j=1}^{r_n}\mathbb{P}(|X_{n,j}|>v_n)\rightarrow0\text{ and }\frac{1}{v_n^2}\sum_{j=1}^{r_n}var(Y_{n,j}|\mathcal{F}_{n,j-1})\rightarrow_p0,$$
	then we have $\frac{1}{v_n}(S_n-B_n)\rightarrow_p0$. Note that the random variables $X_{n,j}$ are not required to have finite expectation. 	In the special case of iid variables, the two conditions become
	$$r_n\mathbb{P}(|X_1|>v_n)\rightarrow 0\text{ and }\frac{r_n}{v_n^2}var(Y_{n,1})\rightarrow0.$$
\end{theorem}

\begin{proof}
 The first condition implies that $\{Y_{n,j}:j=1,\cdots,r_n\}$ are \emph{convergence equivalent} to $\mathbf{X}^{(n)}$ and adapted to $\mathbf{F}^{(n)}$. Put $T_n=\sum_{i=1}^{r_n}Y_{n,j}$ and $T_n-B_n$ is a mean zero martingale adapted to $\mathbf{F}^{(n)}$ so that by conditional Chebyshev's inequality,
 $$U_n=\mathbb{P}(|T_n-B_n|>v_n\epsilon|\mathcal{F}_{n,r_n-1})\le\frac{1}{v_n\epsilon^2}var(T_n-B_n)\rightarrow 0.$$
 By dominated convergence theorem, we have $\mathbb{E}(\lim_nU_n)=\lim_n\mathbb{E}U_n=0$.
\end{proof}

\begin{example}[St Petersburg paradox]
	Nicholas Bernoulli introduced the following game in 1713. A player repeated toss a fair coin until heads appear for the first time. The reward is $2^{n-1}$ dollars if this happends on the $n^{th}$ toss. The question is: what is a reasonable entrance fee (fair stake)  for him to play? The difficulty here is that the expected reward is infinity $(1+1+1+\cdots=\infty)$. Feller's approach was to consider a sequence $\{X_i:i\ge 1\}\}$ as iid random variables such that
	$$\mathbb{P}(X_i=2^n)=\frac{1}{2^n},\ n\ge 1.$$
	Feller defined that the entrance fee $v_n$ for playing $n$ games was fair if $\frac{S_n}{v_n}$ goes to $1$ in probability where $S_n=\sum_{i=1}^nX_n$. His choice of $v_n$ is $n\log_2n$. To prove this is fair, let $k_n=\lceil \log_2v_n\rceil$ so that
	\begin{align*}
		n\mathbb{P}(|X_1|>v_n)&\le\frac{2n}{2^{k_n}}\sim\frac{2}{\log n}\rightarrow 0\\
		\frac{n}{v_n^2}\mathbb{E}(X_1^2\mathbb{I}(|X_1|\le v_n))&\le \frac{n}{v_n^2}\sum_{j=1}^{k_n}2^{2j}\frac{1}{2^{j}}\le \frac{2n}{v_n}\rightarrow0.
	\end{align*}
Hence, by Feller-Kolmogorov we have $\frac{1}{v_n}(S_n-n\mathbb{E}(X_1\mathbb{I}(|X_1|\le v_n)))\rightarrow_p0$. But this means
$$\frac{S_n}{v_n}\sim\frac{n\mathbb{E}(X_1\mathbb{I}(|X_1|\le v_n))}{v_n}\sim\frac{nk_n}{v_n}\sim\frac{k_n}{\log_2n}\rightarrow1.$$
However, as noted in \cite{dabrowska2019}, you should never attempt this with your moneys.
\end{example}
\begin{example}[Exponential spacings and order statistic]
	Let $X_1,\cdots,X_n$ be an iid sample from exponential distribution with mean $1$, and let $X_{(1)},\cdots,X_{(2)}$ be the corresponding vector of order statistics. Further, set $X_{(0)}=0$ a.s. and
	$$T_n=\sum_{i=1}^na_{i}X_{(i)}$$
	where $a_i$ is a sequence of constants. Then using the fact that $Y_j=(n-j+1)(X_{(j)}-X_{(j-1)})$ are iid exponential variables with mean $1$, we have
	\begin{align*}
		\mathbb{E}X_{(k)}&=\sum_{j=1}^k\frac{1}{n-j+1}\\
		cov(X_{(k)},X_{(l)})&=var X_{(k\wedge l)}=\sum_{j=1}^{k\wedge l}\frac{1}{(n-j+1)^2}.
	\end{align*}
Thus, by Chebyshev,
\begin{align*}
	&\mathbb{E}T_n=\sum_{k=1}^n\sum_{i=k}^n\frac{a_i}{n-k+1}\\
	&\mathbb{P}(|T_n-\mathbb{E}(T_n)|>\epsilon v_n)\le\frac{1}{v_n^2\epsilon^2}\sum_{k=1}^n\sum_{j=1}^k\frac{n-k+1}{(n-j+1)^2}.
\end{align*}
Hence, the choices of $B_n$ and $v_n$ in Feller-Kolmogorov can be
\begin{align}
	B_n&=\mathbb{E}T_n\\
	v_n&=\sqrt{n\log n}.
\end{align}
In the special case that $a_i=1,i=1,2,\cdots,$ we have $B_n=n$ and $T_n$ is the sample summation.
\end{example}
\begin{example}[Pareto's distribution]
	Suppose that $X_1,\cdots,X_n$ are iid variables with Pareto distribution and the shape parameter is $1$, i.e., 
	$$\mathbb{P}(X_i>x)=\frac{1}{x}\text{ if }x>1.$$
	Then in this case, $\mathbb{E}X_i$ is infinite. By verifying the conditions in Feller-Kolmogorov,
	\begin{align*}
		n\mathbb{P}(|X_1|>n\log n)&=\frac{1}{\log n}\rightarrow 0\\
		\frac{1}{n\log^2n}\int_1^\infty \mathbb{I}(x\le n\log n)dx\le\frac{1}{\log n}\rightarrow 0.
	\end{align*}
Hence, we have $$\mathbb{P}(|\frac{S_n-n\log n -n\log\log n}{n\log n}|>\epsilon)\rightarrow 0$$
where $\epsilon>0$ and $S_n=\sum_{i=1}^nX_i$. In other words,
$$\frac{S_n}{n\log n}\rightarrow_p1.$$

\end{example}

\subsection{Strong  Consistency} Though subtle difference in practice, the so-called strong consistency has attracted mathematicians and statisticians for decades and it is based on a family of results known as strong law of large numbers (SLLN) \cite{chung2008strong}. One of the most celebrated SLLN in statistics is the \emph{Glivenko-Cantelli theorem}. In fact, it asserts a little more: the strong consistency is uniform. Later the result was extended by Dvoretzky, Kiefer and Wolfowitz in 1956 to get an exponential-type convergence bound \cite{massart1990tight}.
\begin{theorem}[Glivenko-Cantelli \cite{dabrowska2019, wellner2013weak}]
	Let $X_1,\cdots,X_n$ be a sequence of iid sample from an unknown distribution function $F$. Define the \emph{empirical distribution} as
	$$\widehat{F}_n(x)=\frac{1}{n}\sum_{i=1}^n\mathbb{I}(X_i\le x).$$
	Then we have
	\begin{align}
		\mathbb{P}\left(\limsup_{n\rightarrow\infty}\left(\sup_{x\in\mathbb{R}}|\widehat{F}_n(x)-F(x)|\right)\right)=0,
	\end{align}
	i.e., the estimator $\widehat{F}_n(x)$ of the true distribution $F(x)$ is uniformly strong consistent.
\end{theorem}

\section{Asymptotic Normality}

\subsection{Classical CLT}

The CLT used to be the central part of mathematics before 1940s. The definitive answer to CLT is given by William Feller in 1940s. However, the Lindenberg-Lévy CLT is the most well-known and widely-used theorem among statisticians and practitioners. Other versions of CLT such as De Moivre-Laplace CLT and Hajék-Sidak CLT are also used in literature. They have paramount influence in both theory and practice. We provide an example below.
\begin{example}[Kernel density estimator \cite{parzen1962estimation}]
	Let $\{X_i:i\ge 1\}\}$ be iid random variables with density function $f$ w.r.t. the Lebesgue measure. Let
	$$\widehat{f}(x)=\frac{1}{nb_n}\sum_{i=1}^nK(\frac{x-X_i}{b_n})$$
	be a kernel density estimator with bandwidth $b_n$ such that $b_n\rightarrow 0$ and $nb_n\rightarrow\infty$. Suppose that $f$ is twice continuously differentiable and the kernel function $K$ corresponds to a probability density with mean $0$ and finite variance satisfying
	$$\int K^2(r)dr<\infty.$$
	Let $\mathbf{x}=(x_1,\cdots,x_n)^T$ be a vector of $n$ distinct design points, the conclusion is that
	$$\sqrt{nb_n}\left(\widehat{\mathbf{x}}-\mathbb{E}\widehat{f}(\mathbf{x})\right)\rightarrow_d\mathcal{N}(0,D)$$
	where the $j^{th}$ component of $\widehat{f}(\mathbf{x})$ is $\widehat{f}(x_j)$ and $D$ is a diagonal matrix with elements $f(x_j)\int K^2(r)dr$. To see the independence, we compute
	\begin{align*}
		nb_ncov(\widehat{f}(x_i),\widehat{f}(x_j))&=\frac{1}{b_n}\mathbb{E}K(\frac{x_i-X}{b_n}K(\frac{x_j-X}{b_n}))-o(1)\\
		&=\int K(r)K(\frac{x_j-x_i}{b_n}+r)f(x-b_nr)dr-o(1)\\
		&=f(x)\int K(r)K(\frac{x_j-x_i}{b_n}+r)dr-o(1)
	\end{align*}
	which converges to $0$ as $b_n\rightarrow 0$.

	The weak consistency follows from Chebyshev's inequality while the asymptotic normality follows from Lindenberg-Feller CLT. In addition, the strong consistency can be shown via Bernstein's inequality and Borel-Cantelli lemma.
\end{example}

\subsection{Cramer's Conditions}
We assume that $X_1,\cdots,X_n$ are independent random variables that have density w.r.t. some $\sigma$-finite measure $\mu$ belonging to a parametric model $\mathcal{F}=\{f_{i\theta}:\theta\in\Theta,i=1,2,\cdots,n\}$, $\Theta\subset\mathbb{R}^k$. We write $l_{i\theta}(x)=\log f_{i\theta}(x)$. We assume the following Cramer's conditions \cite{dabrowska2019}:
\begin{itemize}
	\item[C1] The parametrization is identifiable.
	\item[C2] $\Theta$ is open.
	\item[C3] The set $A_n=\{x:f_{i\theta}(x)>0,i=1,2,\cdots,n\}$ does not depend on $\theta\in\Theta$.
	\item[C4] For all $x\in A_n$, the functions $l_{i\theta}(x)$ and $f_{i\theta}(x)$ are once Fréchet differentiable w.r.t. $\theta$ and the matrix
	$$\mathcal{I}_n(\theta)=\sum_{i=1}^n\mathbb{E}_\theta(\dot{l}_{i\theta}(X)\dot{l}_{i\theta}(X)^T)$$
	is finite and positive definite at $\theta=\theta_0$. 
	\item[C5] The functions $l_{i\theta}(x)$ and $f_{i\theta}(x)$ are twice Fréchet differentiable at $\theta_0$ for all $x\in A_n$ and differentiation can be passed under the integral sign.
	\item[C6a] The information $\mathcal{I}_n(\theta_0)$ satisfies $\lambda_{\min}(\mathcal{I}_n(\theta_0))\rightarrow\infty$.
	\item[C6b] For any $c>0$,
	$$R_n(c)=\sup_{\theta\in\bar{B}_n(\theta_0,c)}\lVert\mathcal{I}_n(\theta_0)^{-1/2}[\dot{U}_n(\theta)-\dot{U}_n(\theta_0)]\mathcal{I}_n(\theta_0)^{-T/2} \lVert\rightarrow_{\mathbb{P}_{\theta_0}}0$$
	where $\bar{B}_n(\theta_0,c)$ is the closure of
	$${B}_n(\theta_0,c)=\{\theta:\lVert \mathcal{I}_n(\theta_0)^{T/2}(\theta-\theta_0) \lVert<c\}\}$$
	and $U_n(\theta)$ is the first derivative of the log-likelihood function (score function).
	\item[C6c] We have $$\lVert  \mathcal{I}_n(\theta_0)^{-1/2}\dot{U}_n(\theta_0)\mathcal{I}_n(\theta_0)^{-T/2}+I\lVert\rightarrow_{\mathbb{P}_{\theta_0}}0$$
	\item[C7] We have
	$$\mathcal{I}_n(\theta_0)^{-1/2}U_n(\theta_0)\rightarrow_{\mathbb{P}_{\theta_0}}\mathcal{N}(0,I).$$
	That is, the standardized score equation is asymptotically normal.
\end{itemize}
\begin{theorem}[Maximum likelihood estimator]
	Under Cramer's conditions C1-C7, with probability tending to 1, the maximum likelihod estimator $\widehat{\theta}$ exists, is consistent and satisfies the score equation $U_n(\widehat{\theta})=0$. Further,
	\begin{align}
		\mathcal{I}_n(\theta_0)^{T/2}(\widehat{\theta}-\theta_0)\rightarrow_d\mathcal{N}(0,I).
	\end{align}

\end{theorem}

\subsection{Generalized Linear Models (GLM) with Fixed Design}

Let $\{f_\eta:\eta\in\mathcal{H}\}\}$ be a one-parameter canonical exponential family of distributions with open parameter set $\mathcal{H}$. Let
$$l_\eta(x)=\log f_\eta(x)=\eta x-K(\eta)+\log h(x).$$
We assume that $X_1,\cdots,X_n$ are independent variables whose distribution belongs to this family and the parameter $\eta$ depends on fixed covariates, i.e.,
$$\mathbb{E}_{\eta_i}(X_i)=\mu(\eta_i)=\dot{K}(\eta_i)=\psi(\theta^Tz_i)$$
where $\psi:\mathbb{R}\rightarrow\mathcal{H}$ is a known strictly increasing function while $\theta$ is a vector of regression coefficients. Set
$$\eta_i=r(\theta^Tz_i)$$
where $r=\mu^{-1}\circ\psi$ and $\psi^{-1}$ is known as the \emph{link function}. The choice $r(x)=x$ corresponds to the \emph{canonical link} and the model is called the canonical GLM.

\begin{example}[General GLM \cite{cui2022d}]\label{eg:glm}
	The log-density of the $i^{th}$ observation is
	\begin{align}
		l_\theta(x_i,z_i)=r(\theta_0^Tz_i)x_i-K(r(\theta_0^Tz_i))+h(x_i,z_i).
	\end{align}
	If $r$ is a nice differentiable function with range contained in the interior of $\mathcal{H}$, the score equation becomes
	$$U_n(\theta)=\sum_{i=1}^n\dot{l}_\theta(X_i,z_i)=\sum_{i=1}^nz_ir'(\theta^Tz_i)[X_i-\dot{K}(r(\theta^Tz_i))]=0.$$ 
	The Hessian of the log-likelihood is
	$$\dot{U}_n(\theta)=H_n(\theta)=-H_{1n}(\theta)+H_{2n}(\theta)$$
	where
	\begin{align*}
		H_{1n}(\theta)&=\sum_{i=1}^nz_i^{\otimes 2}\ddot{K}(r(\theta_0^Tz_i))[r'(\theta_0^Tz_i)]^2,\\
		H_{2n}(\theta)&=\sum_{i=1}^nz_i^{\otimes 2}[X_i-\dot{K}(r(\theta^Tz_i))]r''(\theta^Tz_i).
	\end{align*}
Taking expectation of the Hessian at point $\theta=\theta_0$, ww find that
$$\mathbb{E}\dot{U}_n(\theta)=-H_{1n}(\theta).$$
Therefore, the term $H_{1n}(\theta)$ reduces to the information matrix $\mathcal{I}_n(\theta)$ for the full model.
\end{example}

 The definitive answer to consistency and asymptotic normality of estimates in GLM is given by Fahrmeier and Kaufman \cite{fahrmeir1985consistency}. We use similar conditions below.

\begin{theorem}[Asymptotic normality in GLM \cite{dabrowska2019}]
	Assume the following regularity conditions:
	\begin{itemize}
		\item[D1] The natural parameter set $\mathcal{H}$ has nonempty interior.
		\item[D2] The function $r$ is a twice continuously differentiable strictly increasing mapping from $\mathbb{R}$ into the interior of $\mathcal{H}$.
		\item[D3] There exist finite constants $C_1<C_2$ such that $[C_1,C_2]$ belongs to the interior of $\mathcal{H}$ and
		$$C_1\le\inf_i r'(\theta_0^Tz_i)\le\sup_ir(\theta_0^Tz_i)\le C_2.$$
		Moreover, there exists a constant $C>0$ such that
		$$\frac{1}{C}\le\inf_i r'(\theta_0^Tz_i)\text{ and }\sup_i\max(r'(\theta_0^Tz_i),|r''(\theta_0^Tz_i)|)\le C.$$
		\item[D4] The information of fixed designs blow up and no dominating design point exists: $\lambda_{\min}(Z_nZ_n^T)\rightarrow\infty$ and $\max_{i\le n}z_i^T(Z_nZ_n^T)z_i\rightarrow0$, where $Z_n$ is the $p\times n$ design matrix.
	\end{itemize}
If $X_1,\cdots,X_n$ are independent and conditions D1-D4 hold, then the maximum likelihood estimator exists with probability tending to 1 and 
\begin{align}
\mathcal{I}_n(\theta_0)^{1/2}(\widehat{\theta}-\theta_0)\rightarrow_d\mathcal{N}(0,I)
\end{align}
where $\mathcal{I}_n(\theta_0)=\sum_{i=1}^nz_i^{\otimes 2}\ddot{K}(r(\theta_0^Tz_i))[r'(\theta_0^Tz_i)]^2$ is the information matrix.
\end{theorem}
\begin{proof}
	To goal is to verify conditions C1-C7 under D1-D4. The conditions C1-C5 are easy to verify based on dominating convergence properties of exponential families. Indeed, the GLM itself and D1 correspond to C1-C2 and D2 implies C4-C5. In the following, we verify conditions C6a-C6c and C7.
	
	Since $K$ forms the cumulant generating function of a full rank exponential family, we have $K^{(p)}(r(\theta_0^Tz_i))$ is finite for all $p$ and $i$ under D1 and first part of D3. Hence, D3 implies that there exists a constant $C$ such that
	$$\mathcal{I}_n(\theta_0)\ge \sum_{i=1}^n\frac{z_i^{\otimes 2}}{C}=\frac{1}{C}Z_nZ_n^T.$$
	Thus, condition D4 implies  C6a. Next, set
	$$a_{i}=\mathcal{I}_n(\theta_0)^{-1/2}z_ir'(\theta_0^Tz_i)\text{ and }b_{i}=\mathcal{I}_n(\theta_0)^{-1/2}z_i$$
	and  by D4 and D5,
	\begin{align*}
	    \max_{i\le n}\lVert a_{i} \lVert^2&\le C^2\max_{i\le n} z_i^T\mathcal{I}_n(\theta_0)^{-1}z_i\rightarrow 0,\\
	    \max_{i\le n}\lVert b_{i}\lVert^2&=\max_{i\le n}z_i^T\mathcal{I}_n(\theta_0)^{-1}z_i\rightarrow 0.
	\end{align*}
\begin{itemize}
	\item We first verify the asymptotic normality condition C7. Similarly, we can WLOG assume there exists a universal constant $C$ such that $\mathbb{E}X_i^4\le C<\infty$ by conditions D1 and D3. We need a version of CLT due to Lyapounov.
	
	\begin{lemma}[Lyapounov CLT \cite{williams1991probability}] Suppose that the variables $X_{ni},i=1,\cdots,r_n$ are independent random vectors with mean $\mu_{nj}$ and non-singular covariance matrix $\Sigma_{nj},j=1,\cdots,r_n$. Define
		$$Y_{nj}=\Sigma_{n}^{-1/2}(X_{nj}-\mu_{nj})$$
		where $\Sigma_{n}=\sum_{i=1}^{r_n}\Sigma_{nj}$. If  $\mathbb{E}\lVert X_{nj}\Vert^{2+\delta}<\infty$ for some $\delta>0$ and 
		$$\lim_n\sum_{j=1}^{r_n}\mathbb{E}\lVert Y_{nj} \lVert^{2+\delta}=0,$$
		then we have
		$$\sum_{j=1}^{r_n}Y_{nj}=\Sigma_{n}^{-1/2}(S_n-\mu_n)\rightarrow\mathcal{N}(0,I)$$
		where $S_n=\sum_{i=1}^{r_n}X_{nj}$ and $\mu_n=\sum_{i=1}^{r_n}\mu_{nj}$.
	\end{lemma}
	Hence, to show condition C7, it is enough to show the Lyapounov's condition holds for $\mathcal{I}_n(\theta_0)^{-1/2}U_n(\theta_0)$. Let $Y_i=X_i-\dot{K}(r(\theta_0^Tz_i))$ so that
	$$\mathcal{I}_n(\theta_0)^{-1/2}U_n(\theta_0)=\sum_{i=1}^na_{i}Y_i.$$
	WLOG, we can assume that $\tau_{4i}=\mathbb{E}Y_i^4\le C<\infty$ for some universal constant $C$. Setting
	$$\sigma_n^2=\sum_{i=1}^n\lVert a_{i}\lVert^2\ddot{K}(r(\theta_0^Tz_i))$$
	so that $\sigma_n^2=\mathbb{E}_{\theta_0}\lVert\mathcal{I}_n^{-1/2}U_n(\theta_0)\lVert^2=p$. Hence,
	$$d_{1}=\sum_{i=1}^n\lVert a_i\lVert^2\le Cp \text{ and }d_2=\sum_{i=1}^n\lVert b_i\lVert^2\le Cp$$
	for some universal constant $C$. The Lyapounov's condition becomes ($\delta=2$)
	$$\sum_{i=1}^n\tau_{4i}\lVert a_i\lVert^4\le C\left(\max_{i\le n}\lVert a_{i}\lVert^2\right)d_1\rightarrow 0.$$
	Hence, we conclude
	$$\mathcal{I}_n(\theta_0)^{-1/2}U_n(\theta_0)\rightarrow_d\mathcal{N}(0,I),$$
	i.e., condition C7 holds.
	
	\item We next verify condition C6c. For $j=1,2,$ put
	$$J_{jn}(\theta)=\mathcal{I}_n(\theta_0)^{-1/2}H_{jn}(\theta)\mathcal{I}_n(\theta_0)^{-T/2}.$$ 
	The matrix $J_{2n}$ decomposes into three parts $J_{2n}=\sum_{k=1}^3J_{2kn}$ where
	\begin{align*}
		J_{21n}(\theta_0)&=\sum_{i=1}^nb_{i}^{\otimes 2}r''(\theta_0^Tz_i)Y_i,\\
		J_{22n}(\theta_0)&=\sum_{i=1}^nb_{i}^{\otimes 2}\left[r''(\theta^Tz_i)-r''(\theta_0^Tz_i)\right]Y_i,\\
		J_{23n}(\theta_0)&=-\sum_{i=1}^nb_i^{\otimes 2}r''(\theta^Tz_i)\left[\dot{K}(r(\theta_0^Tz_i))-\dot{K}(r(\theta^Tz_i))\right]
	\end{align*}
and $Y_i$ are defined as before. After some algebra, we have
$$\mathcal{I}_n(\theta_0)^{-1/2}\dot{U}_n(\theta_0)\mathcal{I}_n(\theta_0)^{-T/2}+I=J_{n21}(\theta_0)$$
and importantly,
\begin{align*}
\lVert var(J_{21n}(\theta_0))\lVert&=\lVert \sum_{i=1}^nb_i^{\otimes 2}\ddot{K}(r(\theta_0^Tz_i))[r''(\theta_0^Tz_i)]^2[b_i^{\otimes 2}]^T \lVert
\\
&\le d_2\left(\max_{i\le n}\lVert b_i\lVert^2\right)\left(\max_{i\le n}\ddot{K}(r(\theta_0^T))[r''(\theta_0^Tz_i)]^2\right).
\end{align*}
	By condition D3-D4 and the fact that $d_2\le Cp$, the above term goes to zero, which verifies condition C6c.
	\item Finally, we verify condition C6b. For any $c>0$ and $\theta\in\bar{B}_n(\theta_0,c)$, we have
	\begin{align*}
		|(\theta-\theta_0)^Tz_i|^2&=|(\theta-\theta_0)^T\mathcal{I}_n(\theta_0)^{1/2}\mathcal{I}_n(\theta_0)^{-1/2}z_i|^2\\
		&\le\lVert\lVert^2\lVert\mathcal{I}_n(\theta_0)^{T/2}(\theta-\theta_0)\lVert^2\lVert\mathcal{I}_n(\theta_0)^{-1/2}z_i\lVert^2\\
		&\le c^2\left(\max_{i\le n}z_i^T\mathcal{I}_n(\theta_0)^{-1}z_i\right)\rightarrow 0.
	\end{align*}
Continuity of the functions $r''$ and $r$ yields
$$t_{j}(c)=\max_{i\le n}\sup_{\theta\in\bar{B}_n(\theta, c)}|h_j(\theta^Tz_i)-h_j(\theta_0^Tz_i)|\rightarrow 0$$
where $h_1(s)=\ddot{K}(r(s))[r'(s)]^2$, $h_2(s)=r''(s)$ and $h_3(s)=\dot{K}(r(s))$. Hence
\begin{align}
	\lVert J_{1n}(\theta)-J_{1n}(\theta_0) \lVert&\le d_2 t_1(c)\rightarrow 0,\\
	\lVert J_{23n}(\theta)-J_{23n}(\theta_0) \lVert&\le d_2t_3(c)\left( \max_{i\le n}|r''(\theta_0^Tz_i)+t_2(c)| \right)\rightarrow 0
\end{align}
uniformly in $\theta\in\bar{B}(\theta_0, c)$. In addition, we also have
$$\mathbb{E}_{\theta_0}\left(\sup_{\theta\in\bar{B}(\theta_0, c)}\lVert  J_{22n}(\theta)-J_{22n}(\theta_0)\lVert\right)\le d_2t_2(c)\max_{i\le n}\mathbb{E}_{\theta_0}|Y_i|\rightarrow 0$$
since the square of the expectation is bounded by the variance of $Y_i$. To sum up, we have shown
\begin{align}
\sup_{\theta\in\bar{B}(\theta_0, c)}\lVert J_{2n}(\theta)-J_{2n}(\theta_0)\lVert\rightarrow_p 0.
\end{align}
Together with condition C6c, the desired condition C6b is verified.
\end{itemize}
\end{proof}


\section{Applications to COVID-19 Data}

We illustrate the asymptotic results by fitting a Poisson regression model to a COVID-19 dataset. The dataset is available online \cite{dong2022johns}. To model the relation between the confirmed COVID-19 cases and the geographical location (e.g., location of the state in U.S.), we shall use three specific columns of the dataset: confirmed cases, latitude and longitude. We assume that the relation forms a Poisson model \cite{hayat2014understanding}. That is, let $Y, x_1,x_2$ be the number of cases, longitude and latitude, respectively, assume
\begin{align}
	\mu(x_1,x_2)&=\mathbb{E}Y,\\
	\log\mu&=\beta_0+\beta_1x_1+\beta_1x_2,
\end{align}
where $\beta_i,i=0,1,2$ are parameters to be estimated. We next define $\beta=(\beta_0,\beta_1,\beta_2)^T$ and $x_i=(1,x_{i1},x_{i2}), i=1,2,\cdots,n$. Back to our example \ref{eg:glm}, we have
\begin{align*}
	\eta_i&=\log\mu_i=\beta^Tx_i,\\
	K(\eta_i)&=\mu_i=\exp(\beta^Tx_i),\\
	\dot{K}(\eta_i)&=\exp(\beta^Tx_i),\\
	\ddot{K}(\eta_i)&=\exp(\beta^Tx_i),
\end{align*}
and $r(s)=s$, $\psi(s)=\exp(s)$. In this case, the information matrix is indeed
$$\mathcal{I}_n(\beta)=\sum_{i=1}^nx_i^{\otimes 2}\ddot{K}(\eta_i)=\sum_{i=1}^nx_i^{\otimes 2}\exp(\beta^Tx_i).$$
Thus, under conditions D1-D4, the maximum likelihood estimator in the Poisson model exists (denote it as $\widehat{\beta}$) and
\begin{align}
	\mathcal{I}_n(\beta)^{-1/2}(\widehat{\beta}-\beta)\rightarrow_d\mathcal{N}(0,I).
\end{align}
The estimator $\widehat{\beta}$ can be solved by Newton-Raphson method \cite{nocedal1999numerical} or simply the "glm" function in R \cite{marschner2018package}. The estimates of $\beta$ are
$$\widehat{\beta}=\left(\begin{matrix}
	1.402\times 10^{1}\\
	3.327\times 10^{-2}\\
	2.315\times 10^{-4}
\end{matrix}\right)$$
and the corresponding covariance matrix due to the asymptotic normality is

\begin{align}
	cov(\widehat{\beta})&=\left(\begin{matrix}
	7.725\times 10^{-8}&-2.021\times 10^{-9} &-1.425\times 10^{-11}\\
	-2.021\times 10^{-9}&  7.927\times 10^{-11} & 1.058\times 10^{-11}\\
	-1.425\times 10^{-11}& 1.058\times 10^{-11}&4.488\times 10^{-12}
	\end{matrix}\right)
\end{align}
and the inference procedure shall be based on both results.

\section*{Acknowledgments}

\bibliographystyle{unsrt}  
\bibliography{references}

\begin{thebibliography}{10}

\bibitem{fisher1925theory}
Ronald~Aylmer Fisher.
\newblock Theory of statistical estimation.
\newblock In {\em Mathematical proceedings of the Cambridge philosophical
  society}, volume~22, pages 700--725. Cambridge University Press, 1925.

\bibitem{chung2001course}
Kai~Lai Chung.
\newblock {\em A course in probability theory}.
\newblock Academic press, 2001.

\bibitem{gnedenko1954limit}
BV~Gnedenko, AN~Kolmogorov, BV~Gnedenko, and AN~Kolmogorov.
\newblock Limit distributions for sums of independent.
\newblock {\em Am. J. Math}, 105, 1954.

\bibitem{kiefer1974general}
Jack Kiefer.
\newblock General equivalence theory for optimum designs (approximate theory).
\newblock {\em The annals of Statistics}, pages 849--879, 1974.

\bibitem{bickel1993efficient}
Peter~J Bickel, Chris~AJ Klaassen, Peter~J Bickel, Ya’acov Ritov, J~Klaassen,
  Jon~A Wellner, and YA'Acov Ritov.
\newblock {\em Efficient and adaptive estimation for semiparametric models},
  volume~4.
\newblock Springer, 1993.

\bibitem{vapnik2015uniform}
Vladimir~N Vapnik and A~Ya Chervonenkis.
\newblock On the uniform convergence of relative frequencies of events to their
  probabilities.
\newblock In {\em Measures of complexity}, pages 11--30. Springer, 2015.

\bibitem{huber2011robust}
Peter~J Huber.
\newblock Robust statistics.
\newblock In {\em International encyclopedia of statistical science}, pages
  1248--1251. Springer, 2011.

\bibitem{andersen2012statistical}
Per~K Andersen, Ornulf Borgan, Richard~D Gill, and Niels Keiding.
\newblock {\em Statistical models based on counting processes}.
\newblock Springer Science \& Business Media, 2012.

\bibitem{hoeffding1994probability}
Wassily Hoeffding.
\newblock Probability inequalities for sums of bounded random variables.
\newblock In {\em The collected works of Wassily Hoeffding}, pages 409--426.
  Springer, 1994.

\bibitem{feller1967introduction}
William Feller.
\newblock {\em An introduction to probability theory and its applications, 3rd
  edition}.
\newblock Wiley, 1968.

\bibitem{dabrowska2019}
Dorota~Maria Dabrowska.
\newblock {\em Advaned probability with elements of real analysis and
  statistics}.
\newblock Unpublished lecture notes at UCLA, 2019.

\bibitem{chung2008strong}
KL~Chung.
\newblock The strong law of large numbers.
\newblock {\em Selected Works of Kai Lai Chung}, pages 145--156, 2008.

\bibitem{massart1990tight}
Pascal Massart.
\newblock The tight constant in the dvoretzky-kiefer-wolfowitz inequality.
\newblock {\em The annals of Probability}, pages 1269--1283, 1990.

\bibitem{wellner2013weak}
Jon Wellner et~al.
\newblock {\em Weak convergence and empirical processes: with applications to
  statistics}.
\newblock Springer Science \& Business Media, 2013.

\bibitem{parzen1962estimation}
Emanuel Parzen.
\newblock On estimation of a probability density function and mode.
\newblock {\em The annals of mathematical statistics}, 33(3):1065--1076, 1962.

\bibitem{cui2022d}
Elvis~Han Cui.
\newblock D-optimal approximate design for binary regression and quantal
  response in toxicology studies.
\newblock {\em arXiv preprint arXiv:2209.13191}, 2022.

\bibitem{fahrmeir1985consistency}
Ludwig Fahrmeir and Heinz Kaufmann.
\newblock Consistency and asymptotic normality of the maximum likelihood
  estimator in generalized linear models.
\newblock {\em The Annals of Statistics}, 13(1):342--368, 1985.

\bibitem{williams1991probability}
David Williams.
\newblock {\em Probability with martingales}.
\newblock Cambridge university press, 1991.

\bibitem{dong2022johns}
Ensheng Dong, Jeremy Ratcliff, Tamara~D Goyea, Aaron Katz, Ryan Lau, Timothy~K
  Ng, Beatrice Garcia, Evan Bolt, Sarah Prata, David Zhang, et~al.
\newblock The johns hopkins university center for systems science and
  engineering covid-19 dashboard: data collection process, challenges faced,
  and lessons learned.
\newblock {\em The Lancet Infectious Diseases}, 2022.

\bibitem{hayat2014understanding}
Matthew~J Hayat and Melinda Higgins.
\newblock Understanding poisson regression.
\newblock {\em Journal of Nursing Education}, 53(4):207--215, 2014.

\bibitem{nocedal1999numerical}
Jorge Nocedal and Stephen~J Wright.
\newblock {\em Numerical optimization}.
\newblock Springer, 1999.

\bibitem{marschner2018package}
Ian Marschner, Mark~W Donoghoe, and Maintainer Mark~W Donoghoe.
\newblock Package ‘glm2’.
\newblock {\em Journal, Vol}, 3(2):12--15, 2018.

\end{thebibliography}

\end{document}